\documentclass[12pt,preprint]{aastex}
\begin{document}
\title{Can high-velocity stars reveal black holes in globular clusters?}
\shorttitle{Globular cluster black holes}

\author{G.A. Drukier \&  C.D. Bailyn}
\affil{Dept. of Astronomy, Yale University, P.O. Box 208101, New Haven, CT, 06520-8101, USA}
\email{drukier, bailyn@astro.yale.edu}

\begin{abstract}
We estimate the number of individual, fast-moving stars observable in globular
clusters under the assumption that the clusters contain massive
central black holes which follow the galactic $M_{\rm BH}-\sigma$
relationship. We find that radial velocity measurements are unlikely
to detect such stars, but that proper motion studies could reveal such
stars, if they exist, in the most likely clusters. Thus, {\it HST} proper motion
studies can test this hypothesis in a few nearby clusters.
\end{abstract}


\section{Introduction} 
It has been suggested \citep{G1}, that globular clusters may contain
central black holes with masses lying along the extension of the
correlation between black hole mass and bulge velocity dispersion as
seen for galaxies \citep{fm00,geb00}. However it has also been
suggested that the observational evidence can be otherwise explained
\citep{b1,b2,d3}.  These difficulties of interpretation are compounded by the
small number of observable stars in the cores of globular clusters.
In contrast to a galactic center, current observations approach the
limit when the uncertainty of an aggregate measurement, such as a
velocity dispersion, cannot be reduced by further observations.  Once
every star has been observed, there is nothing more to be done. (See
Drukier et al. 2003 for a general discussion of issues relating to the
observation and interpretation of velocity dispersions in globular
clusters.)  This is particularly the case for 
central black holes, since there is only a relatively small region
over which the black hole provides the dominant contribution to the
gravitational potential.

Most current observational work relating to the possible presence of
central black holes in globular clusters has been done using radial
velocity measurements to determine the central value and radial
gradient of the velocity dispersion.  In this \it Letter \rm we
explore an alternative approach, namely the use of proper motion
measurements to identify individual stars whose velocities are 
strongly influenced by the presence of a central black hole.  We find
that existing crude models suggest that this approach may be fruitful,
and that the considerable theoretical and observational work that will
be required to apply this approach in practice may thus be justified.
In \S 2, we estimate the number of stars with anomalously high 
velocities caused by a central black hole, and in \S 3 we generate
a list of the clusters with the highest probability of observing
such stars.  We find that for plausible assumptions such stars 
might well be observable with HST in the most favorable clusters.

\section{Estimating the number of high-velocity  stars}
\label{s2}
Consider a globular cluster with central velocity dispersion $\sigma_0$, where
this is the dispersion  outside the radius of
influence, $r_h$, of a black hole with mass $M_\bullet$, given by
\begin{equation}
r_h \equiv {{G M_\bullet} \over {\sigma_0^2}}.
\label{E:rh}
\end{equation}
We define $\sigma_\bullet(r)$ as the
one-dimensional velocity dispersion and $\Sigma(r)$ as the stellar
surface density within $r_h$. Outside $r_h$, the influence of the
black hole is too small to significantly affect stellar orbits.  For
simplicity, we assume that the velocities are distributed as a
Gaussian with dispersion $\sigma_\bullet(r)$.  Now, at a radius $r$,
the fraction of stars which will have velocities greater than some
multiple $k$ of the velocity dispersion $\sigma_0$ is given by
\begin{equation}
f(r) = {{2}\over {\sqrt{2\pi}\sigma_\bullet(r)}} \int_{k\sigma_0}^\infty \exp\left(-{{v^2}\over {2 D \sigma_\bullet^2(r)}}\right) dv,
\label{E:f}
\end{equation}
where $D$ is the dimensionality of the observed velocity (1 for radial
velocities, 2 for proper motions, and 3 for space velocities). While
the upper limit could, in principle, be some multiple of the escape
velocity, this is difficult to define without detailed modeling of the
structure of the whole cluster.  Instead, as the complementary error
function drops off rapidly with increasing argument, we will just use
the infinite upper limit shown. This will lead to a slight
overestimate in the number of expected stars, but this effect is small
compared to  that due to the uncertainty in $\beta$  discussed below. With this
limit, equation~(\ref{E:f}) reduces to
\begin{equation}
f(r) = \sqrt{D}\  {\rm erfc}\left({{k\sigma_0}\over{\sqrt{2D}\sigma_\bullet(r)}}\right).
\end{equation}
The total number of stars with ``significantly'' high velocities
(significant being defined here as $k$ times the velocity dispersion $\sigma_0$
outside $r_{h}$) is then given by
\begin{equation}
N = 2\pi \int_0^{r_h} r \Sigma(r) f(r) dr.
\label{E:N1}
\end{equation}

The only existing dynamical investigation of  the effects of a central
black hole on the structure of a globular cluster is that of
\citet{ck} who used the Fokker-Planck equation to integrate a
steady-state, anisotropic distribution function in the vicinity of a
black hole. They found that the projected velocity dispersion and
surface density profiles could be approximated, for $r<r_h$, by
\begin{eqnarray}
\sigma_\bullet^2(r) &=& \left\{
\begin{array}{ll}
0.4 \sigma_0^2 {{r_h}\over {r}}  & r < 0.4 r_h\\
\sigma_0^2 & 0.4 r_h \le r \le  r_h\\
\end{array}
\label{E:sigma}
\right .,\\
\Sigma(r) &=& \Sigma_0 \left({{r_h}\over {r}}\right)^{0.5}.
\end{eqnarray}
 The two terms in equation~(\ref{E:sigma}) deal with the flattening of the 
velocity dispersion profile near $r_h$. 
Substituting these into equation~(\ref{E:N1}) leads, after a suitable rescaling of the integration variable,  to
\begin{equation}
N=2\pi\Sigma_0 D^2 r_h^2 I_\bullet(k,D),
\label{E:N2}
\end{equation}
where 
\begin{equation}
I_\bullet(k,D) = \int_0^{0.4/D}\sqrt{x}\ {\rm erfc}\left(k\sqrt{x\over 0.8}\right)dx
+{\rm erfc}\left({k\over\sqrt{2D}}\right)\int_{0.4/D}^{1/D}\sqrt{x}dx.
\label{E:I1}
\end{equation}
The integral can be evaluated numerically and gives
 $I_\bullet(3,\{1,2,3\}) = \{1.1, 1.5, 1.5\}\times 10^{-2}$ to two figures. By way of
 comparison, if no black hole is present then, assuming
 $\sigma_\bullet(r)=\sigma_0$ and $\Sigma(r)=\Sigma_0$,
\begin{equation}
I_0(k,D) = \sqrt{D}\ {\rm erfc}\left({k\over\sqrt{2D}}\right)\int_{0}^{1/D}x dx,
\label{E:I2}
\end{equation}
for which $I_0(3,\{1,2,3\}) = \{1.4, 6.0, 8.0\}\times 10^{-3}$, and the expected number of stars is given by equation~(\ref{E:N2}) with $I_0$ in place of $I_\bullet$. 

The galactic black hole mass versus velocity dispersion correlation \citep{fm00,geb00}
is usually given in the form
\begin{equation}
M_\bullet = 10^\alpha \left({{\sigma_0}\over{\sigma_\ast}}\right)^\beta M_\sun,
\label{E:mbh}
\end{equation}
where $\sigma_0$ is a suitable velocity dispersion and $\sigma_\ast =
200$ km s$^{-1}$. The uncertainty in the power-law slope is the
largest source of uncertainty in our estimated number of high-velocity
stars. Recent estimates for $\beta$ (the variation in $\alpha$ is small
since all studies agree for $\sigma_0 \sim \sigma_\ast$) range
from $\beta^L=4.02\pm0.32$ ($\alpha^L=8.13\pm 0.06$) \citep{tremaine} to
$\beta^H=4.65\pm0.48$ ($\alpha^H=8.17\pm0.07$) \citep{mf}. These agree
within their quoted errors, but, on extrapolation to the globular
cluster regime, the use of  $\beta^L$ predicts 20 times as many
high-velocity stars as does the use of $\beta^H$.  We therefore give results for
both these slopes below. The original claim for
globular clusters \citep{G1} used $\beta^L$.

Substituting equations~(\ref{E:rh}) and (\ref{E:mbh}) into
equation~(\ref{E:N2}) we arrive at the following
\begin{equation}
N= 2\pi G^2 10^{2\alpha}\sigma_\ast^{-2\beta}I_\bullet(k,D) D^2  \Sigma_0 \sigma_0^{2(\beta-2)},
\label{E:N2a}
\end{equation}
where $\sigma_0$ is measured in km s$^{-1}$ and $\Sigma_0$ is the
number of {\it measurable} stars per square parsec.  What we can
easily observe is not $\Sigma_0$ but the central surface density,
$\mu_0$, so we need to rewrite $\Sigma_0$ in terms of $\mu_0$ for a
reasonable globular cluster stellar population. For a
defined population of stars, let  $\bar L_*$ be the cluster luminosity per
star in that population in solar units, and let $g_*$ be the fraction of
these stars that are usefully measurable. Then for $\mu_0$ in $V$ magnitude per
square arc second, and taking $M_{V\sun}=4.79$
\begin{equation}
\Sigma_0 = 10^{-0.4(\mu_0-26.37)}{{g_*} {\bar L_*}^{-1}}.
\end{equation}
In this case
\begin{equation}
N=I_\bullet(k,D) D^2 10^{-0.4\mu_0}  g_*\bar L_*^{-1} \hat\alpha\sigma_0^{\hat\beta},
\label{E:N3}
\end{equation} 
where $\hat\alpha^L=2.37\times 10^4$, $\hat\beta^L=4.04$,
$\hat\alpha^H=36.0$, and $\hat\beta^H=5.30$.

For our problem, $\bar L_*$ is determined by the luminosity function
in the core of the cluster in question. It needs to take into account
mass segregation effects and other possible population peculiarities
as might be indicated by, for example, color gradients. The measurable
fraction, $g_*$, is an observational selection effect on the luminosity
function. It will depend on the observational technique to be used,
the distance of the cluster, crowding and so forth. 

Since a complete stellar census is a difficult undertaking, we
estimate $\bar L_*$ as follows. Define our population to be all
the stars brighter than some magnitude, $V_d$ (e.g. the expected
magnitude limit of the observations) in a cluster color-magnitude diagram (CMD). Then,
\begin{equation}
{{\bar L_*}} \approx{{\bar L_b}} + f^{-1}{{\bar L_f}}.
\label{E:lbar}
\end{equation}
 The quantities $\bar L_b$ and $\bar L_f$ are the mean luminosities
for the bright and faint parts of the CMD (divided at $V_d$) and $f$
is the ratio of the number of stars brighter than $V_d$ to the number
of those fainter.  $\bar L_b$ can be estimated directly from the
cluster CMD, for which purpose we use the CMDs in the compilation by
\citet{piotto}.  $\bar L_f$ and $f$ can be estimated from the
corresponding theoretical luminosity function (LF).  We use  those
from the models of \citet{silvestri}. One limitation is that these
LFs only include stars from the hydrogen burning
limit to the tip of the red giant branch, so we must use the CMD to
correct the ratio of the number of stars in each part of the LF,
$f_{\rm LF}$, for the contribution of post-RGB stars. If, brighter
than $V_d$, we see in the CMD $n_R$ stars in the regions covered by
the luminosity function, and $n_B$ more evolved stars elsewhere, then
\begin{equation}
f = f_{\rm LF}\left(1+\frac{n_B}{n_R}\right).
\label{E:nf}
\end{equation}
By substituting equations~(\ref{E:nf}) and (\ref{E:lbar}) into
equation~(\ref{E:N3}), we can now estimate $N$ for any given cluster,
subject to the selection of $g_*$. We proceed to do this in the next section.

\begin{deluxetable}{lrrrrrrr}
\tablecaption{Parameters for the most likely clusters\label{T:top choices} }
\tablewidth{00pt}
\tablehead{\colhead{Cluster}       & 
	\colhead{$\mu_0$\tablenotemark{a}} & 
	\colhead{$\sigma_0$\tablenotemark{b}}  & 
	\colhead{d\tablenotemark{c}} &
	\colhead{$r_h$\tablenotemark{d}}       & 
	\colhead{$\bar L_*$} & 
	\colhead{$N\over N_{\rm M15}^L$} & 
	\colhead{$N\over N_{\rm M15}^H$}}
\startdata
  NGC 6388    &  14.55  &  18.9 & 10.0 & 2.5 & 11.4  &  2.5  &   3.7  \\
  NGC 6441    &  14.99  &  18.0 & 11.7 & 2.0 & 11.3  &  1.4  &   1.9  \\
  NGC 7078    &  14.21  &  14.0 & 10.3 & 1.3 & 11.8  &  1.0  &   1.0  \\
  NGC 6715    &  14.82  &  14.2 & 26.8 & 0.5 & 11.4: &  0.6  &   0.6  \\
  NGC 5139    &  16.77  &  22.0 &  5.3 & 6.5 & 11.4: &  0.6  &   1.1  \\
  NGC \phn104 &  14.43  &  11.5 &  4.5 & 2.1 &  9.8  &  0.4  &   0.3  \\
  NGC 6266    &  15.35  &  14.3 &  6.9 & 2.1 & 10.5  &  0.4  &   0.4  \\
  NGC 2808    &  15.17  &  13.4 &  9.6 & 1.3 & 10.7  &  0.4  &   0.4  \\
  NGC 1851    &  14.15  &  10.4 & 12.1 & 0.6 & 11.2  &  0.3  &   0.2  \\
  NGC 6752    &  15.20  &  12.5 &  4.0 & 2.8 & 11.4: &  0.3  &   0.2  \\
  NGC 6093    &  15.19  &  12.4 & 10.0 & 1.1 & 12.5  &  0.2  &   0.2  \\
  NGC 5824    &  15.08  &  11.6 & 32.0 & 0.3 & 13.7  &  0.2  &   0.1  \\
\enddata

\tablenotetext{a}{Central $V$ surface brightness taken from the 2003
Feb revision of the \citet{harris} compilation.}

\tablenotetext{b}{Central velocity dispersion in km s$^{-1}$ from
\citet{pm93} with the exceptions of NGC 7078 \citep{gerssen1}, NGC 6752
\citep{dbvg}, and NGC 5139 \citep{merritt w cen}.}

\tablenotetext{c}{Distance in kpc from \citet{harris}. }

\tablenotetext{d}{Black hole region of influence as defined by
equation~(\ref{E:rh}) in arc seconds, assuming $\alpha^L$ and $\beta^L$.}

\end{deluxetable}

\section{Best Target Clusters}
\label{s3}
We present in Table~\ref{T:top choices} some relevant numbers for the
12 clusters most likely to show evidence for a central black hole. All
other clusters are estimated to have at least a factor of three fewer
observable high-velocity stars than NGC~5824. The value of $r_h$ in
the fifth column of the table is calculated assuming $\alpha^L$ and
$\beta^L$, and is given in arc seconds for the distance in column
4. Note that while this radius scales as $\sigma_0^{\beta-2}$, the
radii for $\beta^H$ are smaller by a factor of approximately 5.

Our estimates of $\bar L_*$ are given in the sixth column of
Table~\ref{T:top choices}.  The \citet{piotto} CMDs cover 9 of our
listed clusters. For those not listed we have used $\bar L_*=11.4$,
the mean for the other 9. These are marked by a colon after $\bar
L_*$.  We have taken $V_d$ for each cluster such that the division is
at $M_V = 4.5$, a magnitude or so below the turn-off for these
low-metallicity systems. For the luminosity functions we use a very
flat mass-spectral-index $x$ of -0.5, where the Salpeter value is
1.35. Mass segregation is likely to have removed most of the low-mass
stars from the cluster center---many of the mass functions are
actually inverted \citetext{see e.g. \citealp{sosin,de00,ads}}---so
this should be a reasonable approximation. In any case, the
luminosity-function-dependent term $f^{-1}\bar L_f$ in
equation~(\ref{E:lbar}) is roughly 10\% that of $\bar L_b$, so the
estimate of $\bar L_*$ is dominated by the stars in the CMD.

The final two columns give the expected number of stars for the two
estimates of the $M_{\rm BH}-\sigma$ relation relative to the number
expected for M~15 (NGC~7078), the cluster with the best studied core.
Only two clusters, NGC~6388 and NGC~6441, are likely to have more
fast-moving stars than M~15 for the lower slope. NGC~5139
($\omega$~Cen) would have roughly the same number as M~15 using the
higher slope, but the absolute numbers are down by a factor of 24 with
respect to the low slope. To complete our estimate, it remains only to
specify $g_*$ to get the number of fast-moving stars we could
detect. We consider two scenarios: radial velocities and proper
motions.

For radial velocity measurements, $g_*$ is likely to be small, as only
the brightest, least crowded, central stars are suitable. For the case
of M~15, if we compare the number of stars with radial velocities as
compiled by \citet{gerssen1} with the {\it HST} photometric list in
\citet{vdm} then, with $r_h=1\farcs3$, we find that about 4\% of the stars
brighter than $V_d$ have velocities. Within $r_h/2$ this is 7\%, but
within $r_h/4$, where most of the fast-moving stars are to be found,
there are no velocities out of the 10 stars available. Thus, for
radial velocities in M 15, $g_*=0.07$ at best, and we would expect to
see $N_{\rm M15}^L=0.1$ stars, moving with a relative radial velocity $\ge
3\sigma_0$ if a black hole is present. That no such star has been seen
is, therefore, unsurprising and uninformative.  Combining the entire
dozen most-likely clusters yields perhaps one or two fast-moving
stars. $N_{\rm M15}^H=0.004$ stars. The number expected is
effectively zero.

\begin{figure}[t]
\epsscale{0.75}
\plotone{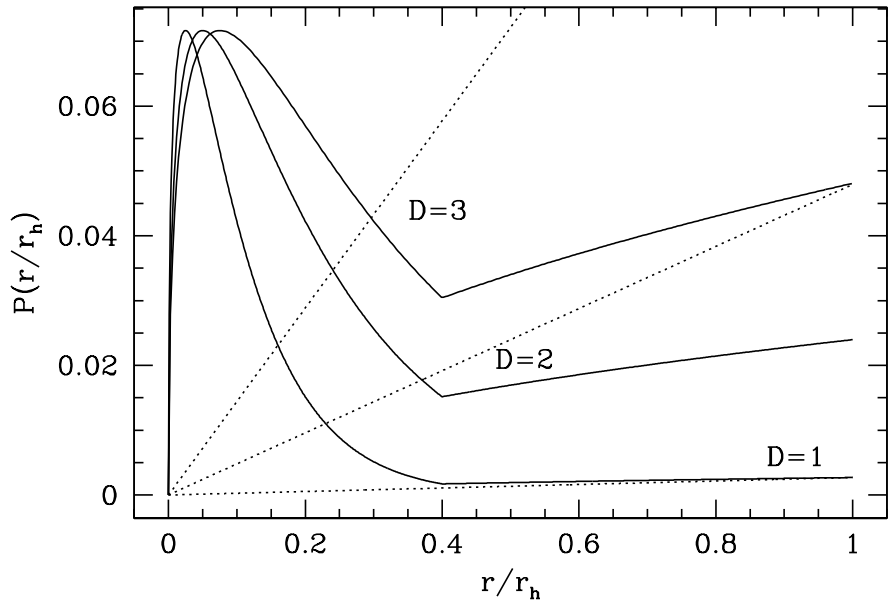}
\caption{The expected radial distribution of high velocity stars with
(solid) and without (dotted) a black hole in the cluster center for
the case $k=3$ and $D=1$, 2, and 3. These distributions are related to
the integrands in equations~(\ref{E:I1}) and equation~(\ref{E:I2})
through a coordinate transformation of the radial variable.  The break
at $0.4r_h$ is due to the break in
equation~(\ref{E:sigma}).\label{F:x}}
\end{figure}

A further problem with using radial velocities to detect the
fast-moving stars lies in their radial distribution as given by the
integrands in equations (\ref{E:I1}) and (\ref{E:I2}) after an
appropriate coordinate transformation. We compare these in
Figure~\ref{F:x} for $k=3$ and the three values of $D$. In all three
cases, the fast-moving stars are concentrated well within $r_h$. The
break point at $0.4r_h$ is the result of the break in $\sigma_\bullet$
at this radius. Note the very different expected radial distributions
for the fast-moving stars depending on whether ({\it solid}) or not
({\it dotted}) there is a black hole. If there is a black hole, the
fast-moving stars should be concentrated in the center of the cluster,
with a maximum inside $0.1 r_h$.   For radial
velocities, the high-velocity stars are confined within about
$0.15r_h$ or 0\farcs 3 for the nearer clusters. For the higher-dimensional
velocities, the detection region is somewhat broader. Clearly, radial
velocities are unlikely to prove the question one way or another.

For proper motions, the situation is much better, at least for nearby
clusters for which such observations are feasible. In the case of NGC
6752, \cite{rb97} measure 153 stars brighter than $V_d$ within
$r_h=2\farcs6$ of the cluster center. To the same radius,
\citet{dbvg}, measure the proper motions of 15 stars. Their detection
region only covers about a third of the area within $r_h$ due to gaps
in their 1999 data. These gaps, unfortunately, covers much of the very
center of the cluster, so it is difficult to estimate the detection
rate within about $0.3r_h$ where we expect to find most of the
fast-moving stars.  Crowding is bound to be a problem in this region,
even for {\it HST}. Extrapolating from the most crowded regions studied
by \citet{dbvg} and taking into account the gaps in their coverage, we
estimate that $g_*=0.2$ is appropriate for proper motions.  In this
case $N_{\rm M15}^L=2$ stars.  If this $M_{\rm BH}-\sigma$ conjecture is
correct, the top three clusters should contain of order 10
high-velocity stars between them with the balance of the 12 clusters
in Table~\ref{T:top choices} contributing another 5 or so in total. In
addition, their radial range is double that for radial velocities. On
the other hand, $N_{\rm M15}^H=0.08$ stars, and the black holes will be
undetectable by this method.

Proper motions have the feature that their uncertainty is inversely
proportional to the time baseline and proportional to the
distance. Nonetheless, the top three clusters, all with distances
between 10 and 12 kpc, should be close enough that, given sufficient
data, the black hole hypothesis can be tested, assuming a value of
$\beta$ at the lower end of its range.  The main limitation at this
distance will be the increase in effective crowding, proportional to
the distance squared.  For the top three clusters in Table~\ref{T:top
choices}, crowding will be 10 to 16 times higher than in the NGC~6752
observations of \citet{dbvg}.

In the absence of a black hole, the flat extrapolation would lead us
to expect to see one or two fast-moving stars within $r_h$ in each of
our top three candidates. These stars will have a radial distribution
proportional to radius, and should be found in the vicinity of $r_h$,
not concentrated within $0.2r_h$ as is predicted under the black hole
conjecture.  Fast-moving stars can have origins other than a
black hole, of course.  Ejection from the core during core-collapse is
one plausible mechanism for producing such stars \citep{dcly}, but
their velocity vectors will be radial unlike a star in orbit around a
black hole.

We caution that the numbers presented here are only estimates and
depend on the scalings found by \citet{ck}. Those models are
single-mass anisotropic Fokker-Planck simulations for the steady-state
stellar distribution in the vicinity of the black hole. More modern
models, which should include, at the very least, a range of stellar
masses and a self-consistent potential, will be needed to fully assess
the significance of any fast-moving stars which are observed in these
clusters. The estimates made here also depend on the current central
velocity dispersions in the globular clusters. Since globular clusters
can lose a large fraction of their mass due to stellar evolution and
stellar-dynamical evolution, the numbers provided in Table~\ref{T:top
choices} may well be underestimated if the proper velocity dispersion
to use in determining black hole masses is the original value, not the
current one.  Further, the mass of any central black hole will have
increased to some extent due to the capture of cluster stars. Using
double the current velocity dispersion, for example, would increase
$N_{\rm M15}^L$ by a factor of 16 and $N_{\rm M15}^H $ by a factor of
39, in which case the higher slope also predicts significant numbers
of observable stars in the top few clusters.  Given the sensitivity to
these effects, obtaining reliable estimates for the numbers of
high-velocity stars will require fully evolving models.  Even in the
event that proper-motion studies uncover no fast moving stars, such
models will allow for firm upper limits on the mass of any black hole.
Constructing such models, and carrying out the necessary proper motion
observations, is no small task, but given the strong interest and
controversy currently surrounding this topic, we believe that efforts
along these lines should be vigorously pursued.

\acknowledgments
This study was supported by a NASA LTSA grant NAG 5-6404.

\clearpage

\clearpage

\begin{thebibliography}{}
\bibitem[Albrow, De Marchi, \& Sahu(2002)]{ads} Albrow, M. D., De Marchi, G., \& Sahu, K. C. 2002, \apj, 579, 660 
\bibitem[Baumgardt et al.(2003a)]{b1} Baumgardt, H.,  Hut, P., Makino, J., McMillan, S., \&  Portegies Zwart, S. 2003a, \apj, 582, L21
\bibitem[Baumgardt et al.(2003b)]{b2} Baumgardt, H., Makino, J., Hut, P., McMillan, \& S., Portegies Zwart, S. 2003b, \apj, 589, L25
\bibitem[Cohn \& Kulsrud(1978)]{ck} Cohn, H.N. \& Kulsrud, R.M. 1978,
\apj, 226, 1087
\bibitem[de Marchi, Paresce, \& Pulone(2000)]{de00} de Marchi, G., Paresce, F., \& Pulone, L. 2000, \apj, 530, 351
\bibitem[Dull et al.(2003)]{d3} Dull, J. D., Cohn, H. N., Lugger,
P. M., Murphy, B. W., Seitzer, P. O., Callanan, P. J., Rutten,
R. G. M., \& Charles, P. A. 2003, \apj, 585, 598
\bibitem[Drukier et al.(1999)]{dcly} Drukier, G.A., Cohn, H.N., Lugger,
P.M., \& Yong, H. 1999, \apj, 518, 233
\bibitem[Drukier et al.(2003)]{dbvg} Drukier, G.A., Bailyn, C.D., Van Altena, W.F., \&  Girard, T.M. 2003, \aj, 125, 2559
\bibitem[Ferrarese \& Merritt(2000)]{fm00} Ferrarese, L. \& Merritt, D. 2000, \apj, 539, L9
\bibitem[Gebhardt et al.(2000)]{geb00} Gebhardt, K. et al. 2000, \apj, 539, L13
\bibitem[Gebhardt, Rich, \& Ho(2002)]{G1} Gebhardt, K., Rich, R.M., \&  Ho, L.C. 2002, \apjl, 578, L41
\bibitem[Gerssen et al.(2002)]{gerssen1} Gerssen, J., van der Marel, R.P., Gebhardt, K., Guhathakurta, P., Peterson, R.C., \&  Pryor, C. 2002, \aj, 124, 3270
\bibitem[Harris(1996)]{harris} Harris, W.E. 1996, \aj, 112, 1487
\bibitem[Merritt \& Ferrarese(2001)]{mf} Merritt, D. \& Ferrarese, L. 2001, in ASP Conf. Ser. 249, The Central Kiloparsec of Starbursts and AGNs, ed. J.H. Knapen, J.K. Beckman, I. Shlosman, \& T.J. Mahoney (San Francisco: ASP), 335
\bibitem[Merritt, Meylan, Mayor(1997)]{merritt w cen} Merritt, D., Meylan, G., \&  Mayor, M. 1997, \aj, 114, 1074 
\bibitem[Piotto et al.(2002)]{piotto} Piotto, G. et al. 2002, \aap, 391, 945
\bibitem[Pryor \& Meylan(1993)]{pm93} Pryor, C. \& Meylan, G. 1993, in ASP Conf. Ser. 50, Structure and Dynamics of Globular Clusters, ed. S.G. Djorgovski \& G. Meylan, (San Francisco: ASP), 357
\bibitem[Rubenstein \& Bailyn(1997)]{rb97} Rubenstein, E.P. \& Bailyn, C.D. 1997, \apj, 474, 701
\bibitem[Silvestri et al.(1998)]{silvestri} Silvestri et al. 1998, \apj, 509, 192
\bibitem[Sosin(1997)]{sosin} Sosin, C. 1997, \aj, 114, 1517 
\bibitem[Tremaine et al.(2002)]{tremaine} Tremaine, S. et al. 2002, \apj, 574, 740
\bibitem[van der Marel et al.(2002)]{vdm} van der Marel, R.P., Gerssen, J., Guhathakurta, P., Peterson, R.C., \&  Gebhardt, K. 2002, \aj, 124, 3255
\end{thebibliography}
\end{document}